# Experimental study of the recombination of a drifting low temperature plasma in the divertor simulator Mistral-B

C. Brault, A. Escarguel, M. Koubiti, R. Stamm,
Th. Pierre, K. Quotb, D. Guyomarc'h

*Laboratoire PIIM, CNRS et Université de Provence, 13397 Marseille, France*

## Abstract

In a new divertor simulator, an ultra-cold ($T_e < 1$ eV) high density recombining magnetized laboratory plasma is studied using probes, spectroscopic measurements, and ultra-fast imaging of spontaneous emission. The Mistral-B device consists in a linear high density magnetized plasma column. The ionizing electrons originate from a large cathode array located in the fringing field of the solenoid. The ionizing electrons are focused in a 3 cm diameter hole at the entrance of the solenoid. The typical plasma density on the axis is close to $2.10^{18}$ m$^{-3}$. The collector is segmented into two plates and a transverse electric field is applied through a potential difference between the plates. The Lorentz force induces the ejection of a very-low temperature plasma jet in the limiter shadow. The characteristic convection time and decay lengths have been obtained with an ultra-fast camera. The study of the atomic physics of the recombining plasma allows to understand the measured decay time and to explain the emission spectra.

## 1. Introduction

A new device has been recently built in order to adress several features of the physics of tokamak divertors. The recombining ultracold plasma is obtained in a way different from the existing divertor simulators for instance PISCES, Nagdis-II or Magnum-PSI [1,2,3]. In our device, an electric extraction of a cold plasma jet is obtained applying a transverse electric field across the central magnetized plasma column. Energetic electrons are axially collected leading to a rapid cooling of the plasma during the early stage of the ExB drift. A low-temperature drifting plasma is formed whose flux is directly related to the central density column and to the applied transverse electric field. At low axial magnetic field, a very fast ion beam is obtained.

## 2. The experimental setup

The new device (see Figure 1) consists in a high pressure cylindrical interaction chamber (140 cm in diameter, 1 meter in length) containing the source plasma. It is connected to a metallic tube (40 cm in diameter, 1.2 meter in length) through an injection hole (3 cm in diameter). The interaction chamber is surrounded by 25 equally spaced water-cooled coils (48 cm internal diameter) producing a magnetic field with intensity up to 0.04 Tesla. The source chamber is located in the fringing field of the solenoid and a thermoionic hot cathode discharge is operated at low pressure. The cathode consists in 32 tungsten filaments (0.2 mm in diameter, 15 cm in lenght) with a heating power of 3 KW. The ionizing electrons are focused on the axis of the target chamber using a magnetic structure inside the source. This multipolar magnetic structure is made of a special arrangement of ferrite permanent magnets.

The working gas (Argon or Helium) is injected at the end of the interaction chamber. Only the source chamber is evacuated through a 1200 l/mn oil diffusion pump. The pressure in the target chamber range from base pressure to 20 Pa.

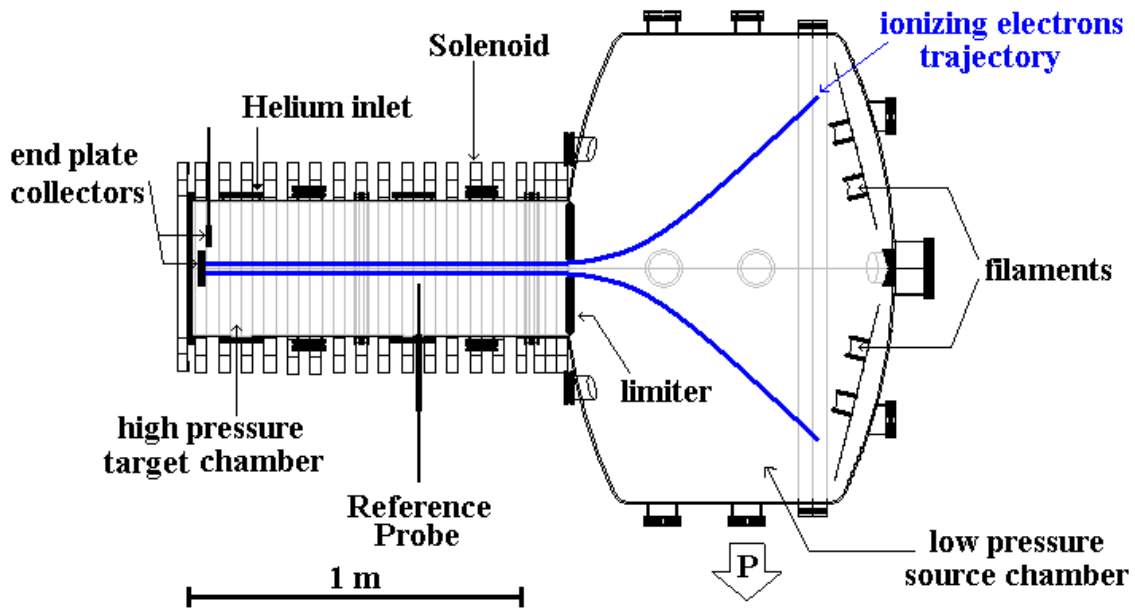

**Figure 1. The Mistral-B device**

The measurement of the plasma parameters are obtained using electric probes, high frequency resonance probes and spectroscopic analysis. The Langmuir probe characteristics allows to measure the radial and axial evolution of the electron temperature. The results are compared to the electron temperature estimated after analysis of the radiation spectra. Using a network analyzer, a UHF double probe with 1 cm radial spacing is used for detection of the lower hybrid resonance cone [4]. This technique gives a good estimate of the obtained electron density inside the central plasma column.

## 3. The ExB extraction

In order to apply a transverse electric field in the central plasma column, the end plate collector is segmented into two plates. The plasma potential across the central plasma column is influenced by the polarisation of the collecting plates. The potential difference between the plates is chosen typically between 5 and 10 volts. This is sufficient to create a transverse electric field along the whole axiell extent of the central column and to obtain the ejection of a plasma jet in the limiter shadow due to the ExB drift of both electrons and ions. The volume of the plasma jet is close to

$10^4$ cm$^{-3}$. This technique has very seldom been used. To the best of our knowledge, this electric extraction has been reported in a numerical study of the efficiency of isotope separation of uranium ions after selective photoionization [5].

It is also possible to used the set-up in pulsed mode applying very short potential variations on the end plates. In this situation, the spatiotemporal evolution of the extracted plasma cloud is tracked using a very sensitive ultra-fast camera developed recently (64 pixels, 16 bits encoding). The characteristic convection time and decay lengths of the recombining plasma can be obtained after analysis of the video record (64.000 frames, 200.000 frames per second).

## 4. Experimental results

The working gas was argon in preliminary experiments in order to test the efficiency of the experimental set-up. The pressure is about 1 Pa inside the target chamber and one order of magnitude lower value inside the source chamber. With an axial magnetic field of 0.02 T, an axial current larger than 20 A is collected by the end plates. In this situation, the estimated density reaches $2.10^{18}$ m$^{-3}$ inside the central magnetized plasma column. The central electron temperature is about 3 eV.

The extraction of the plasma is easily obtained when half of the segmented collecting plate is kept floating. In this case, a transverse radial electric field is induced along the central plasma column. As a consequence, a spiral plasma jet is extracted from the central column. Figure 2 displays the spontaneous light emission of the jet (the central plasma column is not seen behind the end plates. The spatial extension of the jet is related to the recombination processes and to the axial collection of the particles.

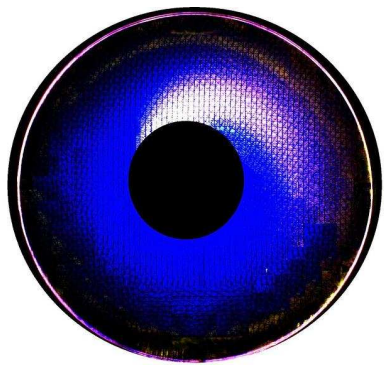

**Figure 2. Plasma jet induced by ExB drift of both electrons and ions**

A radially movable Langmuir probe gives the radial profile of the electron temperature inside the extracted plasma jet. A very fast decrease of the electron temperature is seen inside the jet. The temperature is below 1 eV at a radial position 10 mm from the edge of the central plasma column. This very low value allows for recombination processes to be dominant.

The detailed mechanisms leading to the rapid cooling of the plasma jet are under investigation. First results indicate that the fast ionizing and thermal electrons are lost very quickly by axial collection by the end plates. Behind the limiter, no ionization process occurs and the thermalisation of the extracted plasma jet leads to an equilibrium low electron temperature. It is worth noting that similar mechanisms are present in the SOL of tokamaks and lead to the formation of the electron temperature pedestal. It is also very important to have in mind that in the case of a turbulent central plasma column, transverse turbulent electric fields are self excited inside the column. This leads to the expulsion of plasma bursts around the column in the limiter shadow. Inside the bursts, the same cooling mechanism occurs and leads to the observed recombining turbulent halo around the column. This has been also observed on the Nagdis-II device recently [6] in accordance with our recent results on Mistral [7].

Using helium as working gas, the same results are obtained with a lower value of the collected current. The ultra-cold ejected plasma has been analyzed by emission spectroscopy. A Czerny-Turner spectrometer (50 cm focal length - 1200 mm$^{-1}$ grating) was coupled to a 1024 pixels intensified photodiode array. To improve the signal to noise ratio, very long integration times, between 10 s and 60 s, were applied. The plasma light was coupled to the entrance slit of the spectrometer with an optical fiber and a collimating optic. The line of sight was located along the high pressure source chamber, in the shadow of the limiter. Preliminary spectra were observed in the 335-360 nm spectral region (fig. 2). Emission lines coming from highly excited levels (xd → 1s2p, with x=9..19) can be observed, indicating important radiative recombination processes typical of very low temperature plasmas ($T_e$<1eV). The use of a monochromator coupled to a photomultiplicator device will allow to study the dynamical evolution of the recombining plasma, in correlation with the ultra-fast camera results.

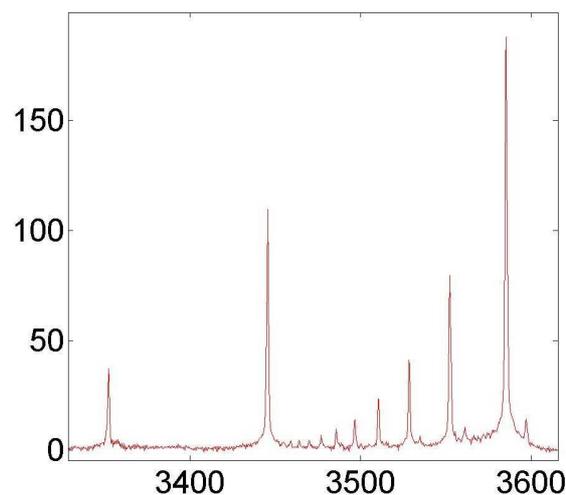

Figure 3 : typical helium spectrum of the recombining ultra-cold ejected plasma.
Recombination lines (9d to 19d → 1s2p) can be observed in the 345 nm-359 nm spectral region.

The camera is located at a distance of 3 meters on the axis of the device. It is imaging a squared section (12x12 cm$^{2)}$) of the target chamber with 1.5 cm spatial resolution. The high sensitivity of the camera is obtained using photodetectors with a amplification factor and 16 bits A/D parallel conversion of the signals. Each record is up to 64.000 frames duration at a record rate of 200.000 frames per second. When the central plasma column is turbulent, bursts of plasma are convected around the column by ExB drift. In this situation, the record exhibits luminescent spots. The trajectory of each spot is clearly seen and the spatiotemporal evolution is analyzed. Figure 4 depicts a series of pictures inside one event with 5 µs time lag between each frame. The convection velocity is inferred from the frames giving typically a few km/s as ExB drift. This corresponds to a transverse electric field in the range 50 to 100 V/m. The local measurement of the density inside the jet gives an estimate for the ion flux across the B-field in the range $10^{19}$-$10^{20}$ m$^{-2}$.s$^{-1}$.

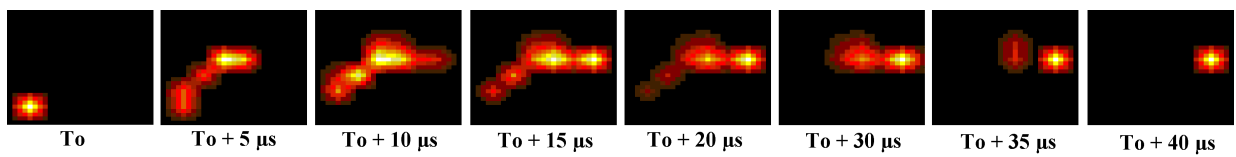

To   To + 5 µs   To + 10 µs   To + 15 µs   To + 20 µs   To + 30 µs   To + 35 µs   To + 40 µs

**Figure 3. Ultra fast record of the expulsion of one the plasma bursts (200,000 frames/sec)**

## 5. Conclusion

We have performed experiments in a new divertor simulator where an ultra-cold recombining plasma is obtained by ExB extraction. The electron temperature is measured using spectroscopic and probe measurements. Typical recombination spectra in helium are recorded. A large volume of recombining plasma is obtained. The spatiotemporal evolution of the recombining plasma is analyzed using an ultra-fast camera.

### Aknowledgments

The authors thanks especially Dr. G. Leclert for very helpful discussions, and G. Vinçonneau and P. Martinez for technical support.